\documentclass[%
 reprint,
 amsmath,amssymb,
 aps,
]{revtex4-2}

\usepackage{graphicx}
\usepackage{dcolumn}
\usepackage{bm}
\usepackage{xcolor}
\usepackage{subfiles}
\usepackage{wrapfig}
\usepackage{xr}
\newcommand{\SK}[1]{\textcolor{black}{{#1}}}

\begin{document}

\preprint{APS/123-QED}

\title{\SK{Inverse Bauschinger Effect in Active Ultrastable Glasses}}

\author{Rashmi Priya$^1$} 
\author{Smarajit Karmakar$^1$} 
 \email{smarajit@tifrh.res.in}
\affiliation{ $^1$ Tata Institute of Fundamental Research, 36/P, Gopanpally Village, Serilingampally Mandal,
Ranga Reddy District, Hyderabad, India 500046}

\begin{abstract}

\SK{Memory effects in amorphous materials have been widely studied because of their possible widespread future applications. We show here that ultrastable glasses can exhibit a transient reversible memory effect when subjected to both a local driving force via Run-and-tumble active particles and global shear. We investigate the system's response across different yielding regimes by selectively switching the shear direction at different strains. We analyze how changes in shear direction influence yielding, post-yield behavior, and structural evolution in active amorphous solids. Our model active system exhibits an enhanced anisotropic response, displaying both conventional and inverse Bauschinger effects, depending on the deformation history. The results indicate that activity-induced shear band networks create structural memory, enabling the system to heal upon shear reversal due to the transient nature of this phenomenon. Additionally, we observe that shear softening under cyclic loading produces an irreversible, stable, and less branched network structure with increasing cycles. These findings provide novel insights into how activity and shear collectively contribute to mechanical response, including memory formation in ultrastable disordered systems.}

\end{abstract}

\maketitle


\noindent{\label{sec:level1}\bf \large Introduction:\protect}
\SK{Memory effects are tied inherently to systems that are far from equilibrium. In equilibrium, the initial conditions are forgotten, and no memory persists. Therefore, studying different types of memory in far-from-equilibrium systems provides deeper insights into the system's dynamical evolution. Various forms of material memory, including unidirectional shear memory, Mullins effects, Kovacs effects, return-point memory, memory of maximum amplitude from oscillatory driving, and shape memory, as well as ageing and rejuvenation in glasses \cite{Bauschinger, DIANI2009601, Kovac, bouchbinder2010nonequilibrium, PhysRevLett.107.010603, PhysRevLett.112.025702, Adhikari2018-eg, PhysRevLett.113.068301, PhysRevLett.104.257201, PhysRevLett.81.3243, RevModPhys.91.035002} have been extensively studied in the literature. Among these, the memory of the direction is one of the most fundamental properties exhibited by amorphous materials in the form of the Bauschinger effect. This effect, widely studied by \cite{Bauschinger, PhysRevResearch.6.033211, PhysRevLett.124.205503, PhysRevE.82.026104}, reflects the system's ability to "remember" the direction of prior shear deformation. When an amorphous material or even a polycrystalline metal is subjected to unidirectional shear, its microstructure develops anisotropy due to the imposed stress. Upon reversing the shear direction, the material exhibits a reduced yield stress compared to the original loading curve, revealing an asymmetry in its mechanical response. This effect is a direct consequence of plastic rearrangements that imprint the shear history onto the system’s microstructure, creating a mechanical memory that influences future deformation and serves as a quantifiable signature of past loading conditions \cite{RevModPhys.91.035002}.}

\SK{Ultrastable glasses are brittle and lack plastic drops in the elastic regime \cite{https://doi.org/10.1002/adma.201302700, PhysRevLett.125.168003, PhysRevLett.125.085505, 2022NCimR..45..325R} making them generally not a good candidate for storing mechanical memory. Once a shear band forms, the system undergoes irreversible plastic deformation, consistently relaxing through the same pathway and leading to permanent structural changes. However, with the introduction of activity, we demonstrate that the system's behavior changes significantly. The emerging network-like structure facilitates smaller plastic drops after yielding, and the material remains less plasticized, as we have observed in our concurrent work \cite{rpriya2025}. We have also found that activity can enhance mechanical stability by promoting gradual failure. In this study, we utilize the gradual formation of multiple shear bands and show that this process retains a degree of reversibility if negative shear is applied during the transient stage of network development. As the network continues to evolve, reversibility becomes increasingly difficult, eventually saturating at a specific stress value in the steady state. We further demonstrate that the network structure plays a crucial role in governing this behavior. In \cite{PhysRevMaterials.4.025603}, the emergence of shear band network structure has been found at higher strain rates in passive ultrastable glasses.} 

\SK{In this work, we have found a stronger asymmetry between forward (positive) and reverse (negative) shear in active amorphous solids as compared to passive systems when starting from a pre-sheared state, suggesting that after unloading, the material retains a memory of the prior shear direction. In the passive case, this effect is well-known as the Bauschinger effect, where the system resists yielding in the same direction but yields more easily in the other orthogonal directions. However, with activity, we observe a more intricate response: initially, the system yields more easily in the prior direction but resists yielding in the opposite direction. This behavior, referred to as the inverse Bauschinger effect, reflects anisotropic mechanical memory and the healing of transient shear bands in the presence of activity.} 

\SK{The inverse Bauschinger effect is a fascinating phenomenon that is known to happen in certain materials where the yield strength increases when the direction of loading is reversed. This counterintuitive behaviour is argued to happen in crystalline solids under deformation due to mechanisms that constrain the movement of dislocations, which are defects in the crystal structure, from moving back in the opposite direction \cite{PhysRevResearch.6.033211}. Multiple studies have identified several factors that contribute to this effect, including the presence of deformation twins, which restrain dislocation movement, and strain hardening, where continued plastic deformation enhances the material's strength even when the loading direction changes \cite{ZHOU202215}. Grain boundaries are also believed to play a significant role, as dislocation interactions with these grain boundaries can further inhibit their movement \cite{ZHOU202215, Koizumi2016}. Notably, materials such as single- or polycrystalline aluminium and high-strength steels often exhibit this inverse Bauschinger effect under specific conditions \cite{Koizumi2016, Lopes2009}. We believe that our observation of the inverse Bauschinger effect in amorphous solids could be the first documented observation of the inverse Bauschinger effect in amorphous solids. Therefore, a deeper understanding of this phenomenon is essential for accurately predicting how materials will behave under complex loading scenarios, particularly in applications that involve cyclic loading and active particles.} 

\SK{The healing of shear bands under cyclic shear, when a strain amplitude lower than the yield strain is applied, has been previously studied \cite{PhysRevX.9.021018}. In this work, we report the healing of the transient shear band network under homogeneous shear upon reversal of the shear direction. Due to this reversible nature of shear band formation under shear reversal, the shear band network delays yielding and requires greater deformation before failure occurs. This is a significant finding, demonstrating that shearing in one direction strengthens the system against deformation in the opposite direction. Finally, as the system transitions to a steady state, permanent shear band formation results in a response that converges toward the passive case. Notably, a denser shear band network retains a broader range of residual stress and structural memory upon shear reversal, making it more suitable for studying such effects.}

\SK{The mean orientation of a particle's contact normal provides valuable insight into its structural anisotropy. This has been widely studied in granular and amorphous solids using the fabric tensor \cite{radjai2004key, PhysRevLett.102.195501, PhysRevLett.124.205503}. In silica glass, \cite{PhysRevLett.102.195501} shows that amorphous silica develops permanent anisotropy when subjected to external shear. Similarly, our study also observes this behavior. The anisotropy parameter derived from the fabric tensor can effectively capture structural anisotropy and helps in analyzing how shear and activity modify the microstructure of ultrastable glasses.}   

\SK{Response to cyclic shear deformation in amorphous solids recently gained lot of attention in the scientific community as cyclic shear enables glassy systems to reach lower energy configurations that are otherwise inaccessible through conventional annealing protocols. When subjected to repeated cyclic shear, non-equilibrium systems exhibit mechanical annealing at small strain amplitudes (below yield strain) and form shear bands at larger strain amplitudes (above yield strain), which is associated with energy dissipation and loss of memory of the initial condition \cite{10.1063/5.0100523, PhysRevE.88.020301, PhysRevLett.124.225502, Leishangthem2017-ya}. In conventional ultrastable 2D glassy systems, no mechanical annealing is observed in the elastic regime, and the energy remains nearly constant. However, when the strain amplitude exceeds the yield point, a discontinuous jump occurs with the formation of a shear band \cite{PhysRevLett.124.225502, 10.1063/5.0085064}. In contrast, when activity (at lower persistence time) is introduced, we observe a distinct post-yield behavior: after yielding, the stress response progressively decreases with each cycle until reaching steady-state saturation.} This behavior, distinct from that of passive ultrastable glasses, appears to be previously unexplored and opens up new avenues for exploring Bauschinger effect in amorphous solids using a combination of active forcing and oscillatory shear.
 
\SK{This phenomenon of progressive decrease in yield stress, known as shear softening, suggests that the system effectively learns to flow more easily under repeated active perturbations, reducing its ability to store and recover stress efficiently in post-yielded active ultrastable glasses.} Cyclic shear gradually reshapes the material's mechanical memory by promoting the formation of well-connected, deeply plasticized, system-spanning shear band networks. This leads to a progressive reduction in mechanical resistance and initiates a transition toward a steady-state as cycles continue, where classical Bauschinger effects are expected to emerge. Such behavior is particularly common in soft glassy materials, where cyclic loading beyond the yield point results in strain localization and formation of shear bands \cite{PhysRevX.9.021018, JANA2020120098}. The material effectively adapts to the deformation pathway, facilitating structural rearrangements and internal stress relaxation. 

\SK{The rest of the article is organised as follows. First, we briefly discuss the models and the simulation details, and how we introduce activity in this model system, and then in the result section, we first discuss the stress response of active amorphous solids under stress reversal and characterise the development of anisotropy in the system. Then we discuss the Bauschinger effect in the presence of activity, especially the inverse Bauschinger effect at small deformation and recovery of the classical Bauschinger effect at large deformations. We then discuss the response of the system under oscillatory shear deformation and highlight the evolution dynamics of shear bands in the presence of activity.}

\section{Models and Methods \protect\\}

\noindent{\bf \large Model:}
\SK{We model a system consisting of polydisperse particles interacting via a soft-sphere potential. The mean particle diameter $\bar \sigma$ is set to $1.0$, with the maximum and minimum diameters being $1.61 \bar \sigma$ and $0.725 \bar \sigma$, respectively, such that $\sigma_{max}/\sigma_{min} = 2.22$ as proposed in \cite{PhysRevX.7.021039}. The interaction rule between particle diameters is non-additive and is defined as $\sigma_{ij} = \frac{\sigma_i + \sigma_j}{2} (1-0.2 |\sigma_i - \sigma_j| )$, which enhances the structural stability. Particle $i$ interacts with particle $j$, separated by a distance $r_{ij}$, via the following potential with a cutoff distance of $r_c = 1.25$:
\begin{eqnarray}
V(r)=
\begin{cases}
r^{-12} + C_0 + C_2 r^2 + C_4 r^4 &  ,r < r_c \\
0   & ,r \geq r_c
\end{cases}
\label{Potential}
\end{eqnarray}
where $r = r_{ij}/\sigma_{ij}$. The coefficients are chosen as $C_0 = -28/r_c^{12}$, $C_2 = 48/r_C^{14}$, and $C_4=-21/r_c^{16}$ so that the potential and its derivatives are continuous at the cutoff.}

\vskip +0.1in
\noindent{\bf \large Simple Shear Simulations and Activity:}\SK{To shear the 2D system, we first prepared the system at $T=0.026$ using molecular dynamics (MD) with swap Monte Carlo (SMC) \cite{Berthier_2019}. The prepared sample was then quenched to a lower temperature of $T=0.001$. Then, it was sheared using the SLLOD equations of motion with a Gaussian thermostat \cite{10.1063/1.479358, 10.1063/1.1858861, 10.1093/oso/9780198803195.001.0001}. To perform the same process in the presence of the activity, we first prepared the ultrastable glass, quenched it, and then introduced activity and shear. An active force $f_0$ was applied to $20\%$ of particles, chosen randomly. The active particles mimic the run-and-tumble motion, running in one direction for a persistent period $\tau_p$ before switching direction while conserving net momentum. The active force modifies the equation of motion as follows:
\begin{eqnarray}
\boldsymbol{\dot{r}} &=& \frac{\boldsymbol{p}_i}{m}\\
\boldsymbol{\dot{p}}_i &=& -\frac{\partial \boldsymbol{V}}{\partial \boldsymbol{r}}_i+n_i \boldsymbol{f}_i,
\label{EOM}
\end{eqnarray}
where $n_i = 1$ for active particles and $0$ otherwise. $\vec f_0$ is defined as follows: 
\begin{eqnarray}
  \vec{f_0}=f_0(k_x \hat{x} +k_y \hat{y} ), 
  \label{Active}
\end{eqnarray}
where $k_x$ and $k_y$ control the direction in which any active particle moves. The values of $k_x$ and $k_y$ are chosen from $\pm 1$, which applies the force along four diagonal directions, similar to the 4-state clock model. The sum of $k$ in all directions is maintained at zero to ensure momentum conservation. After every $\tau_p$, the direction of the active particles is randomized to mimic the run-and-tumble motion observed in various biological and synthetic active particles.} 

\vskip +0.1in
\noindent{\bf \large Shear Band Analysis:} To analyze shear bands, we compute non-affine displacements using the $D^2_{min}$ \cite{PhysRevE.57.7192}. This quantity measures local deviations from an affine deformation field, capturing plastic rearrangements in amorphous solids and glassy systems under shear. An interaction range of $2.5$ is used to identify the neighbors of each particle, which are then used to calculate the local strain tensor and minimize the affine part of the motion. The resulting $D^2_{min}$ values at any strain are calculated with respect to the unstrained system and are used to characterize shear bands. These values typically range from $0.0$ to $1.0$ and are colour-coded accordingly unless otherwise specified. A value of $0.0$ indicates regions outside the shear bands, while a value of $1.0$ represents areas with the highest non-affine displacement, typically lying inside the shear band.

\begin{figure*}
\includegraphics[width=0.99\textwidth]{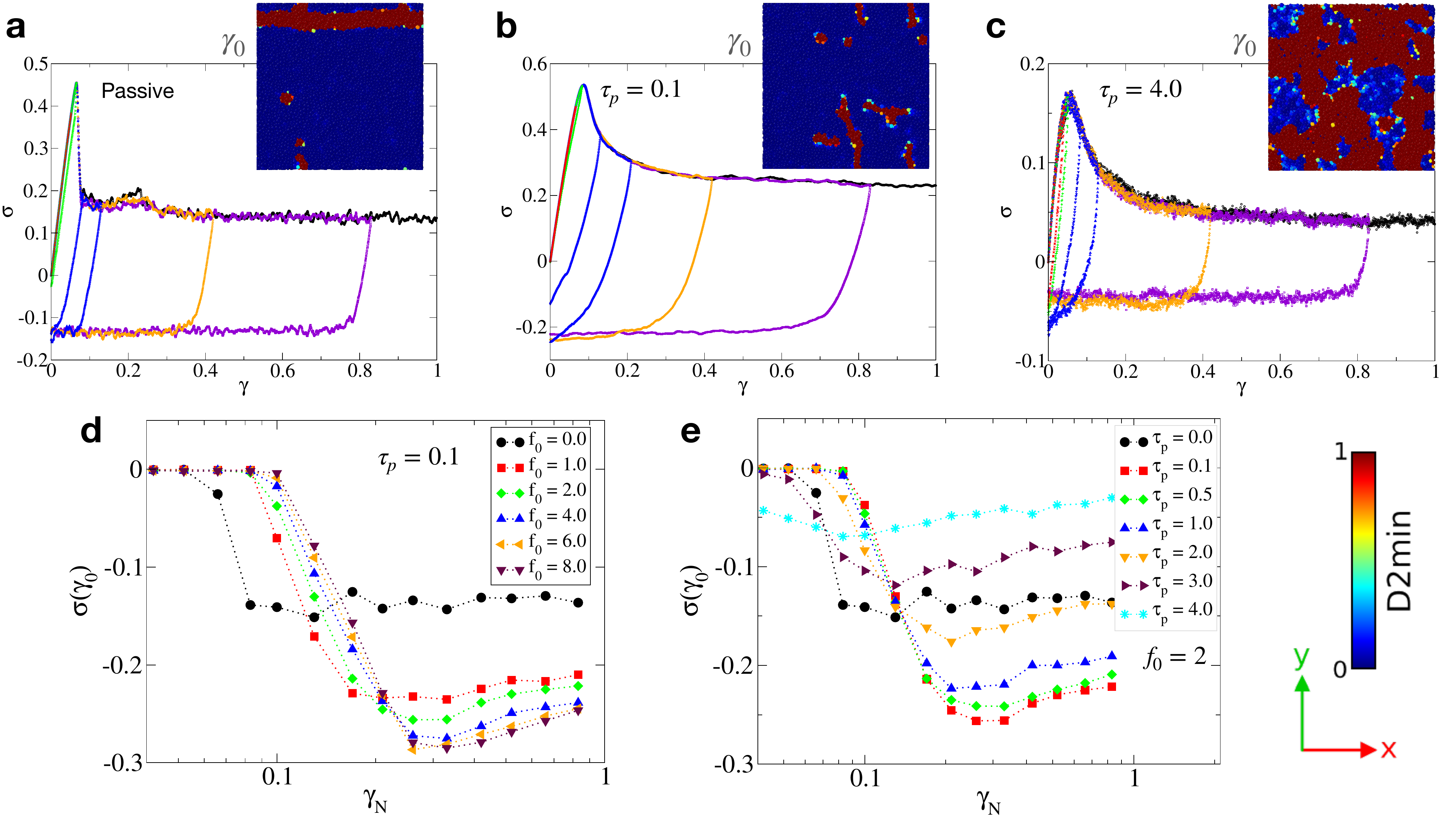}
\caption{{\bf Stress Response Under Strain Reversal: }  Effect of switching strain direction at different strain values in passive and active systems. (a) The passive system and (b) low persistence time $\tau_p =0.1$ and $f_0=2.0$ system show no hysteresis in the elastic regime (red and green curves). After crossing the yield point, the passive system drops to negative stress immediately due to shear band formation, while the active system with $\tau_p = 0.1$ exhibits a gradual decrease in residual stress (blue curves). This behavior is evident in panel (d), where the stress at zero strain is plotted against maximum strain $\gamma_N$. For the passive case, this stress remains nearly constant after yielding. In contrast, for active systems with low $\tau_p$ across all studied $f_0$ values, the stress decreases with increasing strain after yielding, then rises slowly before saturating as the system approaches steady state. For higher $\tau_p = 4.0$ and $f_0=2.0$ shown in (c), hysteresis and stress recovery are similar, but increased mobility and formation of smaller mobile regions at smaller strains lead to small hysteresis effects even before yielding. Shear band images at zero strain ($\gamma_0$) after strain reversal show nonzero stress due to the presence of mobile regions, as seen in the snapshots in (a–c). The stress at zero strain for different $\tau_p$ values at $f_0 = 2.0$, along with the passive case, is shown in (e). All cases with lower persistence time ($\tau
_p \leq 2.0$) exhibit a gradual decrease, followed by a slow increase after a certain strain value.}
\label{BE1}
\end{figure*}

\vskip +0.1in
\noindent{\bf \large Shear Reversal Protocols:}
\SK{To study the memory effect in the ultrastable glass, the system was subjected to shear deformation in both directions, with or without activity. After shearing to a maximum strain value $\gamma_N$, we reversed the direction of shear following the path: $\gamma_{0} \rightarrow \gamma_N \rightarrow \gamma_0$, where $\gamma_0 = 0.0$ is the zero strain. We selected values of $\gamma_N$ using the formula $\gamma_N = \gamma_Y \times 10^{\pm 0.1i}$, where $i$ is an integer ranging from $[0,n]$, constrained such that $ 0.0 <\gamma_N < 1.0$. Here, $\gamma_Y$ represents the yield strain corresponding to the passive system with the strain rate $\dot{\gamma} = 5 \times 10^{-5}$. The stress-strain curve identifies two noteworthy states: the zero-stress state ($\gamma(\sigma_0)$, $\sigma_0$) and the zero-strain state ($\gamma_0$, $\sigma(\gamma_0)$). Using the zero-stress state as our initial condition, we investigated the Bauschinger effect by shearing the system in both positive and negative directions. We then compared these responses with those obtained when freshly prepared states were sheared in positive and negative directions. We also studied the system's response to oscillatory shear with $n<5$ cycles. For this investigation, we strain the system following the path $\gamma_{max} \rightarrow 0 \rightarrow -\gamma_{max} \rightarrow 0 \rightarrow \gamma_{max}$ for both passive systems and active systems.}

\vskip +0.1in
\noindent{\bf \large Anisotropy Characterization - Fabric Tensor:}
\SK{Fabric tensors measure the microstructural anisotropy of a system by analyzing the directions of contact normals between interacting particles. The eigenvalues and eigenvectors of the fabric tensor provide insight into the anisotropy of the system. The eigenvalues and eigenvectors of the fabric tensor are computed per particle and then averaged over the system. 
In a perfectly isotropic system in 2D, the two eigenvalues of the normalized fabric tensor are $1/2$, and their sum equals $1$. As anisotropy increases, the averaged eigenvalues become different, with the maximum eigenvalue and the corresponding eigenvector indicating the dominant contact orientation. To quantify this, a scalar quantity known as the anisotropy index ($\alpha$) is defined:
\begin{eqnarray}
 \alpha = \frac{1}{N}\sum_{i=1}^N{\alpha_i} 
\end{eqnarray}
where $\alpha_i$ is anisotropy index for individual particles defined as: 
\begin{eqnarray}
 \alpha_i = 2 \sqrt{ \sum_{j=1}^2{(\lambda_{j} - \frac{1}{2})^2}} 
\end{eqnarray}
where $\lambda_{j}$ are the eigen values for $i^{th}$ particle in the x and y directions and they satisfy $\sum_j \lambda_j =1$. In an isotropic system, the anisotropy index is $\alpha = 0$. For anisotropic systems, $\alpha > 0$ indicates deviation from isotropy, and in the ideal scenario of complete anisotropy, the index reaches its maximum value of $\alpha = 1.0$. In our analysis, we account for the system’s polydispersity when defining the cutoff for contact normals.}

\begin{figure}
\includegraphics[width=0.48\textwidth]{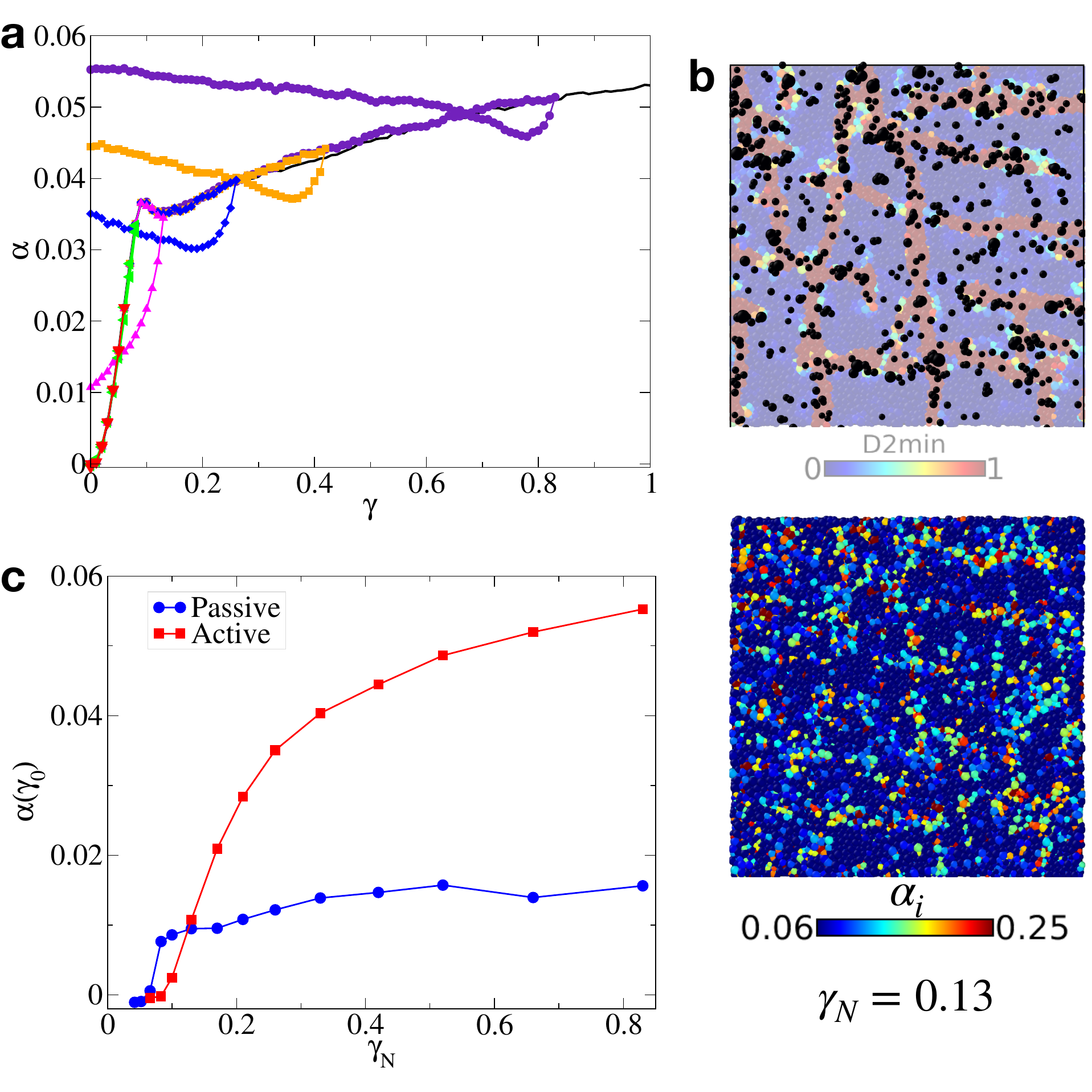}
\caption{{\bf Anisotropy Index:} The evolution of anisotropy index during loading-unloading for various deformation histories ($\gamma_N$ equals 0.06 (red), 0.08 (green), 0.13 (pink), 0.26 (blue), 0.42 (orange), 0.83 (indigo)) is shown in panel (a) for an active system with $\tau_p = 0.1$ and $f_0 = 2.0$. In (b), the spatial correlation between structural anisotropy and plasticity is shown at strain $\gamma = 0.13$ for a maximum loading of $\gamma_N = 0.13$. The top panel overlays particles with a high anisotropy index ($\alpha_i > 0.25$, shown in black) on non-affine displacements ($D^2_{min}$), which identifies plastically active zones. This overlay reveals a strong spatial correlation between regions of significant contact network anisotropy and the shear band network. The bottom panel of (b) depicts the spatial distribution of the anisotropy index, with values ranging from $0.06$ (blue) to $0.25$ (red), illustrating that directional bias in interparticle contacts is concentrated along the pathways of plastic deformation. Finally, panel (c) compares the residual anisotropy index ($\alpha(\gamma_0)$) after unloading to zero strain between active and passive systems across different loading histories.}
\label{AnisotropyIndex}
\end{figure} 

\section{Memory in Post-Yielded Ultrastable Glass}

\noindent{\bf \large Stress Response Under Strain Reversal:}
\begin{figure*}
\includegraphics[width=0.99\textwidth]{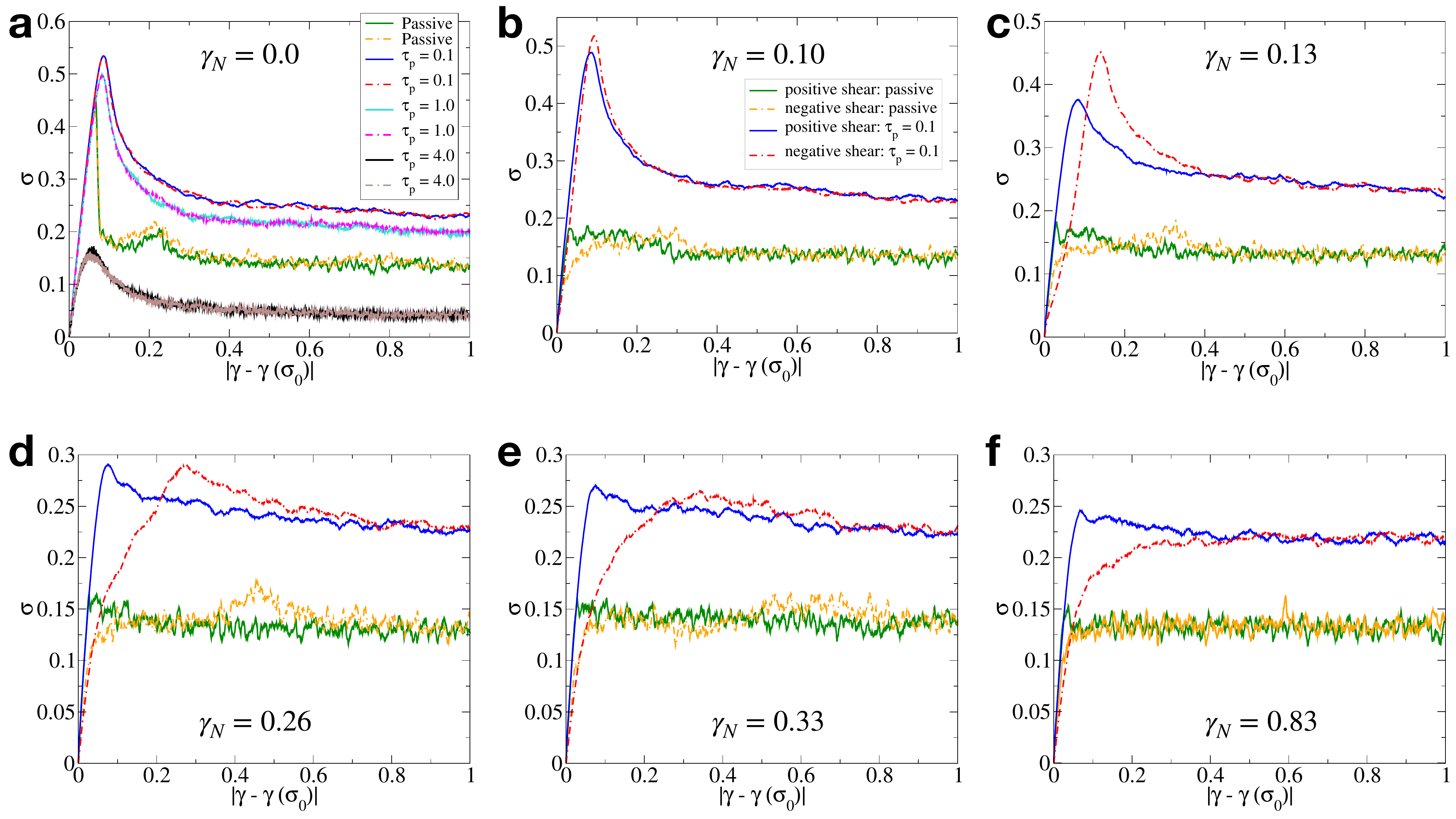}
\caption{{\bf Transition from Inverse to Classical Bauschinger Effect in Active Solids:} In Fig. (a), we demonstrate the stress-strain response of the freshly prepared system, which exhibits symmetric behavior when subjected to identical strain rates in both positive and negative directions. This symmetry is consistently observed across passive systems and active systems with varying persistence times ($\tau_p$) at a fixed active force magnitude $f_0 = 2.0$. Figs. (b-f) illustrate the system's response when sheared from a pre-shear state with zero stress upon unloading. The strain values $\gamma_N$ represent the maximum strain up to which the system was initially sheared before unloading to the zero-stress state. The system demonstrates a pronounced anisotropic response between forward and reverse shear directions, with this anisotropy evolving systematically as $\gamma_N$ increases. At lower $\gamma_N$ values, the stress magnitude in the reverse direction surpasses that in the forward direction, accompanied by an increase in yield strain, which is the inverse of the classical Bauschinger effect. However, as $\gamma_N$ increases, the stress during reverse shearing progressively decreases, and as the system approaches steady-state, it ultimately exhibits the classical Bauschinger effect.}
\label{BE2}
\end{figure*}
\SK{Traditional way to identify unidirectional memory effect in any system is to subject it to shear in a reverse direction after shearing the system once in a forward direction and then analyze the relative difference in its stress-strain response. In Fig.~\ref{BE1}, we examine the effect of reversing the strain direction at different strain values $\gamma_N$. To distinguish this effect from the conventional response of ultrastable glass, we conducted simulations for different persistence times and active forces and compared them to the passive system. Fig.~\ref{BE1}(a) presents the mechanical response of a passive system sheared at a strain rate of $\dot{\gamma} = 5 \times 10^{-5}$. In the elastic regime, the stress-strain curve for negative shear follows the same path as the forward shear. However, the system exhibits an immediate stress drop beyond the yield point at $\gamma=0.08$. A residual negative stress emerges at zero strain when the shear direction is reversed just after yielding. As strain increases further, the stress saturates around the same value. This behavior arises due to the formation of a persistent shear band at zero strain, $\gamma_0$, as illustrated in the corresponding snapshot. The mobility of regions within this shear band is responsible for the observed negative stress in the passive system.}  

\SK{Fig.~\ref{BE1} (b) and (c) present the stress response for systems with persistence times of $\tau_p = 0.1$ and $4.0$, respectively, at different values of $\gamma_N$. The reversal of direction in the elastic regime does not cause any hysteresis, as indicated by the red and green curves but beyond the yield point system no longer follows the same path; but decreases and goes to negative stress whose value keeps decreasing as shown in blue color and then starts increasing very slowly and saturates at higher $\gamma_N$. A detailed analysis of stress at zero strain $\sigma_(\gamma_0)$, for passive and active systems with $\tau_p=0.1$ at different active force magnitudes, $f_0$ is shown in Fig.~\ref{BE1} (d).} {Higher active forces at lower persistence times result in lower residual stress at zero strain for $\gamma_N > 0.2$. The lower stress also implies that the non-affine regions of the shear band present at zero strain have also decreased (possible healing of the shear band). This reduction in stress suggests a decrease in the non-affine regions of the shear band at zero strain. 
Thus, we show that the presence of activity is crucial for obtaining a lower $\sigma(\gamma_0)$ and facilitating its gradual decrease.} 

\SK{In panel (c), the effect of hysteresis and the decrease followed by an increase in stress at zero strain are similar. Figure~\ref{BE1}(e) illustrates the effect of varying persistence times on stress evolution with increasing $\gamma_N$ at $f_0 = 2.0$. Note that with increasing $\tau_p$ beyond $0.1$, the stress at zero strain for large $\gamma_N$ shows a systematic increase, eventually becoming greater than the passive case at large enough $\tau_p$. Also, larger $\tau_p$ results in smaller mobile regions at very small strains; the stress-strain response does not follow the same path and exhibits a hysteresis loop even before yielding. The shear band images shown for all cases are at zero strain, showing non-zero non-affine displacement quantified using $D^2_{min}$. This analysis reveals that the residual stress originates from the presence of a shear band or mobile regions in all three cases.}

\vskip +0.1in
\noindent{\bf \large Quantifying Anisotropy - Fabric Tensor Analysis:}
\SK{To quantify the anisotropy developed in the system during the loading process, we computed the fabric tensor derived anisotropy index for an active system with $\tau_p = 0.1$ and $f_0 = 2.0$ under different loading-unloading protocols characterized by maximum strain $\gamma_N$ as shown in Fig. ~\ref{AnisotropyIndex} (a).}{ For maximum strains below the yield point $\gamma_N < \gamma_Y$ ($\gamma_N = 0.06, 0.08$), the system recovers its original state upon unloading. However, as loading increases beyond yield ($\gamma_N = 0.13, 0.26, 0.42, 0.83$), the contact network develops persistent anisotropy that remains even after unloading to $\gamma = 0$.} \SK{This residual anisotropy in the fabric tensor emerges from the formation of shear band networks, as illustrated in Fig. ~\ref{AnisotropyIndex} (b), where particles with high anisotropy index values ($\alpha_i > 0.25$) are superimposed on the $D^2_{min}$ field that identifies plastically active sheared zones at strain $\gamma = 0.13$. The bottom image shows the spatial distribution of the anisotropy index within a range of $0.06$ to $0.25$ for clarity, revealing that while there is spatial heterogeneity, the highest anisotropy values strongly correlate with regions of plastic deformation. The patterns in both panels show clear similarities, suggesting that regions undergoing plastic deformation develop stronger contact anisotropy. This suggests that microscopic structural changes persist even after unloading.} 

\SK{Fig. ~\ref{AnisotropyIndex}(c) compares the anisotropy index between passive and active systems, demonstrating that active systems develop significantly more pronounced contact anisotropy, which increases systematically with loading history. This enhanced anisotropic response in active systems stems from their more extensive and gradually evolving shear band networks, demonstrating the fundamental differences in how activity affects microstructural evolution under mechanical deformation and the overall yielding process in active solids.} 

\begin{figure*}
\includegraphics[width=0.99\textwidth]{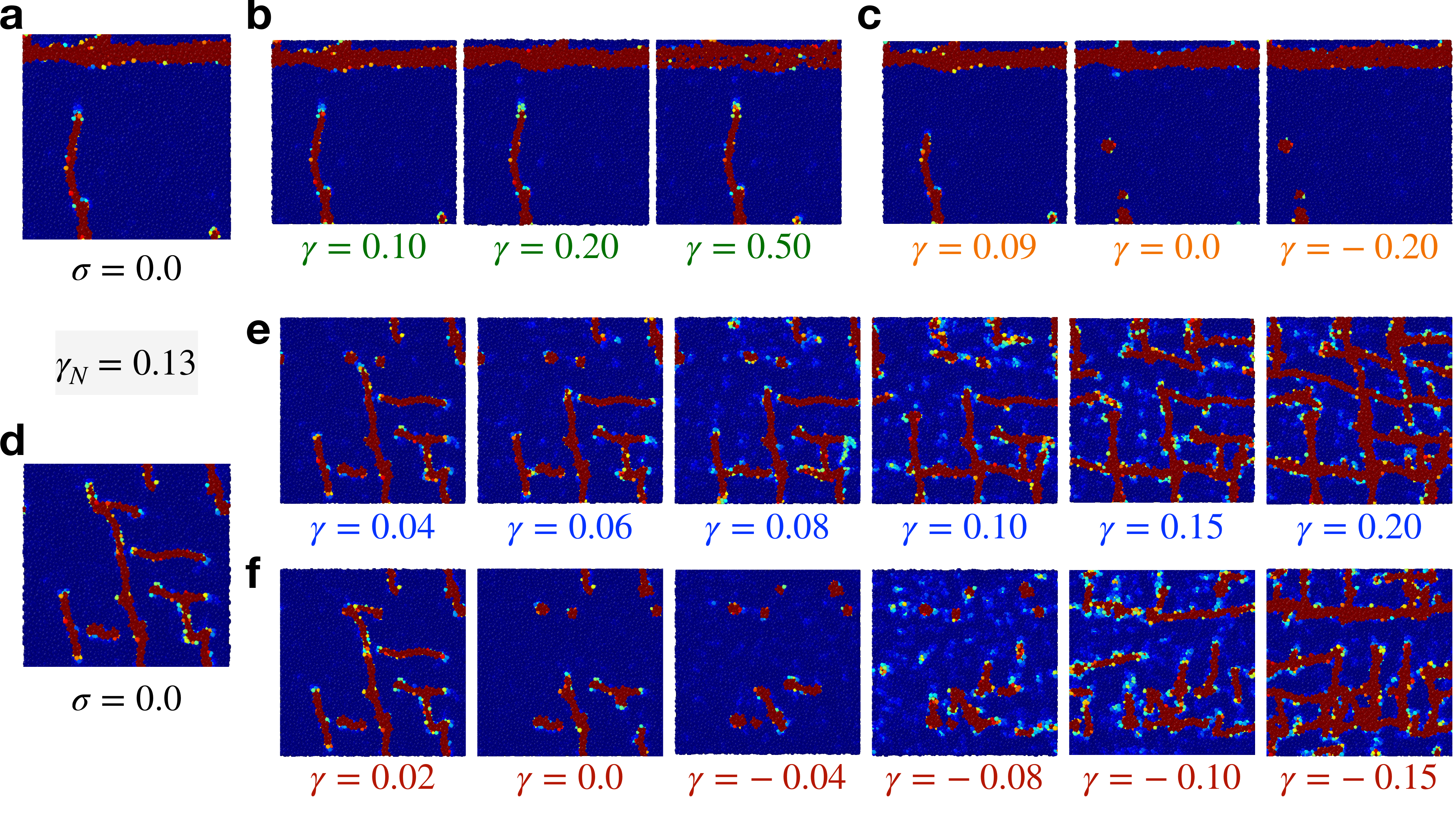}
\caption{{\bf Healing and Reformation of Shear Bands}: This figure illustrates the evolution of shear bands in passive (a-c) and active (d-f) systems subjected to shear in the positive (green/blue) and negative (orange/red) directions. The color scheme is consistent with that used in the stress–strain curves in Fig.~\ref{BE2} for clarity. The maximum strain at which the direction of shear was reversed to reach a zero-stress state in both passive and active cases is $\gamma_N = 0.13$, and the corresponding stress–strain response from this zero-stress state is shown in Fig.~\ref{BE2} (c). Panels (a) and (d) show the initial zero-stress states, with respective strain values of approximately $0.095$ for the passive case and approximately $0.024$ for the active case, from which shearing begins in both directions. Panels (b) and (e) depict shear band formation in the forward direction, while panels (c) and (f) illustrate shearing in the negative direction. In the passive case, the system-spanning shear band persists in the reverse direction, although a secondary, weaker, and incomplete band fades. This effect is more pronounced in the active case with $\tau_p = 0.1$ and $f_0 = 2.0$. Since no fully developed shear bands are present in panel (d), forward shearing leads to shear band development between strains of $0.06$ and $0.10$, resulting in early network formation. Conversely, in the negative direction, we observe healing of the shear band, where red regions (shear bands) gradually vanish between $0.02$ and $-0.06$. Around $\gamma = -0.08$, new shear bands begin to form at the sites of small red regions that serve as weak spots, eventually developing into a fully evolved network as the strain progresses. This phenomenon explains the delayed yielding observed in the reverse direction of the stress-strain curve.}
\label{BE2SB}
\end{figure*} 

\vskip +0.1in
\noindent{\bf \large Transition from Inverse to Classical Bauschinger Effect :}
\SK{To identify the remnants of structural deformation, we compare the mechanical response of the system beyond yielding in two cases: freshly prepared samples subjected to shear and samples pre-loaded to a strain $\gamma_N$ and subsequently unloaded to a zero-stress state. This analysis is conducted for different deformation histories, $\gamma_N$, to assess how prior loading influences the mechanical response. Figure.~\ref{BE2} (a) presents the mechanical response for freshly prepared samples, while panels (b-f) show the responses of systems with prior loading history. The stress-strain response remains symmetric for samples without any strain history in both passive and active systems, with different persistence times for an active force of $f_0=2.0$. The dotted line refers to the shear in reverse directions, while solid lines represent shear in forward directions. However, when strain history is introduced, asymmetry emerges. In the active system (red-blue), the response at $\gamma_N = 0.10$ (panel b) shows minimal asymmetry, with the system exhibiting greater resistance to yielding in the reverse direction. This effect becomes more pronounced at $\gamma_N = 0.13$, where an inverse Bauschinger effect is observed. The yield stress (stress peak) increases by a large amount when the shear direction is reversed. As $\gamma_N$ increases, the yield strain continues to increase while the yield stress gradually decreases. Around $\gamma_N = 0.26$, the yield stresses in both forward and reverse shear become equal. Beyond this point, the system transitions to the classical Bauschinger effect, where the yield stress in the reverse direction decreases, a phenomenon well-documented \cite{}. The yield stress in both shear directions remains lower than that of the original stress-strain curve due to the presence of remnant stress regions, which act as weak spots that facilitate shear band formation. In contrast, the passive system (green-orange) exhibits the classical Bauschinger effect for all strain histories.}  

\vskip +0.1in
\noindent{\bf \large Healing of the Shear Band Networks: }
\SK{The evolution of shear band networks for $\gamma_N = 0.13$ is shown in Fig.~\ref{BE2SB} and is directly linked to the observed inverse Bauschinger effect in Fig.~\ref{BE2}(c). Panels (a–c) depict the passive system, where (a) represents the initial zero-stress state, (b) shows the forward shear, and (c) illustrates the reverse shear. The strain levels mentioned below each figure panel are with respect to the strain ($\gamma_0$) at which the bulk stress in the system reaches zero while performing the shear reversal from the strain value $ \gamma_N = 0.13$. The primary shear band remains intact, but smaller mobile regions disappear during reverse shear for the passive case. In the active system (d–f), panel (d) shows the initial zero-stress state. During forward shear (e), the shear band rapidly develops from mobile regions, forming an interconnected network. In the reverse shear process (f), the system initially heals the shear band at small strains ($0.02$ to $-0.06$). However, some remnant regions with high non-affine displacements persist, acting as weak spots that serve as nucleation sites for subsequent deformation, leading to the reformation of the shear band network.} These weak spots contribute to a lower yield stress in both directions compared to the freshly prepared system and are responsible for shear softening at large deformations. 

\SK{With the increase in the reversal strain value, $\gamma_N$, the shear band becomes more deeply plasticized, making healing progressively more difficult. The system thus retains a memory of prior deformation, leading to a decrease in reverse yield stress, similar to passive systems. Different pre-shear strains $\gamma_N$ dictate the stage of shear band network formation, governing the system's anisotropic mechanical response. This tunability between inverse and classical Bauschinger effects demonstrates how the introduction of activity can modulate the system's ability to retain or erase structural deformations, offering insight into how active materials can possibly alter mechanical memory in ultrastable glasses in a fundamental manner.}

\vskip +0.1in
\noindent{\bf \large Structural Reorganization and Shear Softening under Oscillatory Shear: }
\begin{figure}
\includegraphics[width=0.48\textwidth]{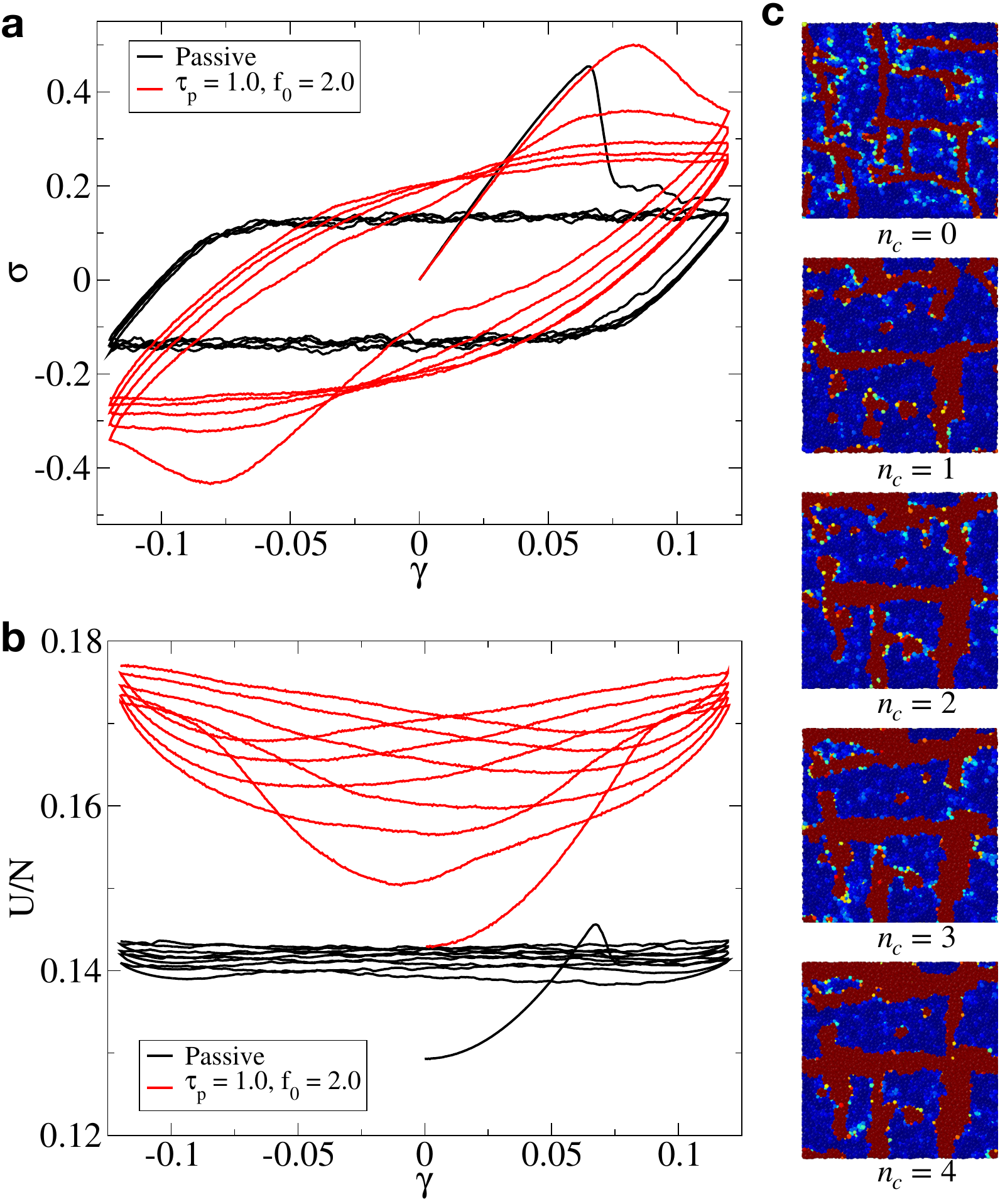}
\caption{{\bf Structural Reorganization and Shear Softening:} This figure presents the stress-strain response in (a) and potential energy with strain response in (b) under oscillatory shear, performed at $\gamma_{max} = 0.12$. In the passive system (black curves), we observe minimal changes in both the stress-strain and potential energy-strain curves. The system follows the same loop across cycles, indicating that it retains its original mechanical state, i.e., the same shear band after each cycle, which means it does not undergo significant structural change. In contrast, the active solid with $\tau_p=1.0 $ and $f_0=2.0$ undergoes progressive shear softening, with a gradual decrease in yield stress in both directions over successive cycles. This implies the material becomes easier to deform as cycling continues. Additionally, the potential energy per particle increases continuously, indicating that the system evolves into a mechanically softer state with more mobile regions. In (c), we see that with an increase in the number of cycles, the snapshots of shear bands taken at $\gamma_{max} = 0.12$ for each cycle show that particles reform in new configurations after yielding to reach a lower stress state. This happens in the process of healing weak networks and formation of stronger ones, which is irreversible.}
\label{OscShear}
\end{figure}
\SK{Healing of shear band networks provides a direct means to erase structural memory, emphasizing that the system’s deformation history strongly governs memory formation and retention. Cyclic shear provides another systematic approach to study the influence of strain history on mechanical behavior through repeated loading cycles of $\gamma_{max} \rightarrow 0 \rightarrow -\gamma_{max} \rightarrow 0 \rightarrow \gamma_{max}$, as shown in Fig.~\ref{OscShear}. In a passive system, shearing at a strain rate of $\dot{\gamma} = 5 \times 10^{-5}$, a hysteresis in the stress-strain curve is observed. The potential energy of the system increases only minimally, as the material does not undergo significant structural rearrangements. The stress-strain and potential energy curves follow a reproducible cyclic loop, indicating that the system retains its mechanical memory. In contrast, the active system at $\tau_P=1.0$ and $f_0 = 2.0$ exhibits a progressively decreasing yield stress with each cycle, known as shear softening. This means that less stress is required to induce plastic flow with every cycle.}

\SK{This behavior arises from accumulated plasticity, where irreversible deformations lead to a less rigid structure. Initially, a dense shear band network forms and governs deformation, as shown in Fig.~\ref{OscShear}(c) at $n_c = 0$. Upon shear reversal, this network partially vanishes before a new, deeper, and less dense network emerges with increasing strain. Particles reorganize into configurations that facilitate lower stress states, as observed at $n_c = 1$. This process involves the healing of weaker structures and the formation of persistent networks due to irreversible microstructural changes. The disappearing red pathways (shear banded regions) in the blue matrix (non-banded regions) indicate where previously stress-bearing networks have collapsed or reformed. Meanwhile, emerging networks show where particles have found more favorable arrangements to distribute the applied loads. With each cycle, the newly formed shear band becomes increasingly plasticized, progressively reducing the influence of the prior band. Due to accumulated plastic work and increased disorder, potential energy rises.} However, once the system approaches a steady state, the energy and the stress-strain response gradually converge as observed in a passive system, with cycle-to-cycle differences diminishing progressively (Supplementary Fig. 2).  


\SK{Interestingly, while active particles initially hinder the formation of a system-spanning shear band, cyclic shear ultimately promotes its emergence through repeated structural reorganization and easier deformation/flow.} The evolution of shear band networks suggests that ultrastable glasses, under cyclic shear, evolve into a mechanically softer and energetically less stable state, characterized by enhanced flow and persistent plastic deformation. Note that with increasing number of oscillations, the shear band structure evolves to a steady configuration and the system exhibits the classical Bauschinger effect once more. This reaffirms once again that the formation and evolution of shear bands play a key role in governing the observed inverse Bauschinger effect in active solids.   

\section{Discussion}
\SK{Our findings in active amorphous solids align with and extend known memory effects in amorphous materials, particularly the Bauschinger effect, the inverse Bauschinger effect, and memory encoding through plastic rearrangements. After unloading, we observe nonzero remnant stress and anisotropy in active ultrastable glasses, indicating history-dependent microstructural changes similar to those reported in amorphous silica \cite{PhysRevLett.102.195501}. The spatial correlation between non-affine displacements and the anisotropy index suggests that structural changes arise from particle rearrangements driven by stress relaxation.} 

\SK{Additionally, we demonstrate that activity is crucial for reducing remnant stress at zero strain after unloading, $\sigma(\gamma_0)$, and facilitating its gradual decrease. The interplay between activity parameters ($\tau_p$ and $f_0$) and strain history ($\gamma_N$) governs the system’s ability to retain structural deformations and shapes its stress response. Thus, activity enhances the system’s capacity to retain or erase memory. Without activity, a system that fails via shear band formation undergoes minimal structural change upon strain reversal, exhibiting the classic Bauschinger effect. In contrast, activity promotes gradual shear band formation, leading to an inverse Bauschinger effect, which transitions to the conventional Bauschinger effect as strain progresses. The same phenomenon is observed if one does oscillatory shear deformation cycles. Under multiple shear cycles, the shear band network evolves into a more permanent network, which does not show any further restructuring (both creation and healing of shear band branches), and the system transitions to a state where one observes the classical Bauschinger effects. This suggests that the healing and restructuring of multiple shear band networks play a central role in the inverse Bauschinger effect in active solids. It will be worth exploring further whether such a mechanism is also at play even for passive solids, where one observes inverse Bauschinger effects along with the other known effects, such as dislocation twins, which prevent free movement of dislocations in crystalline solids.}


\SK{Finally, our results highlight the crucial role of activity in shaping the mechanical memory of ultrastable glasses by regulating shear band formation and evolution. Using the characteristics of networked shear bands, one can offer a framework for quantifying shear history and mechanical memory in far-from-equilibrium systems. The interplay between activity and deformation history determines whether structural deformations are retained or healed, providing a framework for tuning the mechanical response of disordered solids through controlled pre-shear. Future studies could explore the possibilities of storing and retrieving the mechanical memory in active ultrastable glasses. The ability to encode and erase memory through activity and cyclic shear post-yielding may present a pathway for controlling memory retention in ultrastable glasses.} 

\vskip +0.05in
\begin{acknowledgments}
\SK{We acknowledge funding from intramural funds at TIFR Hyderabad, provided by the Department of Atomic Energy (DAE) under Project Identification No. RTI 4007. SK acknowledges Swarna Jayanti Fellowship grants DST/SJF/PSA01/2018-19 and SB/SFJ/2019-20/05 from the Science and Engineering Research Board (SERB) and Department of Science and Technology (DST). Most computations are performed using the HPC clusters procured through Swarna Jayanti Fellowship grants DST/SJF/PSA01/2018-19 and SB/SFJ/2019-20/05. SK would like to acknowledge the research support from the MATRICES Grant MTR/2023/000079, funded by SERB. SK also thanks Kunimasa Miyazaki for the interesting discussions and warmly acknowledges his hospitality during SK's visit to Nagoya University, where part of the manuscript was written. SK also acknowledges travel support from the JSPS invitational fellowship.}
\end{acknowledgments}

\bibliography{activeYieldingUltraStable}

\end{document}


\title{\textbf{Supplementary Information - Inverse Bauschinger Effect in Active Ultrastable Glasses}}
\author{Rashmi Priya$^1$} 
\author{Smarajit Karmakar$^1$} 
\affiliation{ $^1$ Tata Institute of Fundamental Research, 36/P, Gopanpally Village, Serilingampally Mandal,
Ranga Reddy District, Hyderabad, India 50046}
\maketitle

Movie Files are added to show the healing and reformation of shear bands for $\gamma_N = 0.13$ and $0.17$ under forward and reverse shear, starting from the zero stress state for the active system at a strain rate of $\dot{\gamma} = 5 \times 10^{-5}$, persistence time $\tau_p = 0.1$ and active force magnitude of $f_0=2.0$.

\section{Stress Response Under Strain Reversal}

\begin{figure*}
\includegraphics[width=0.99\textwidth]{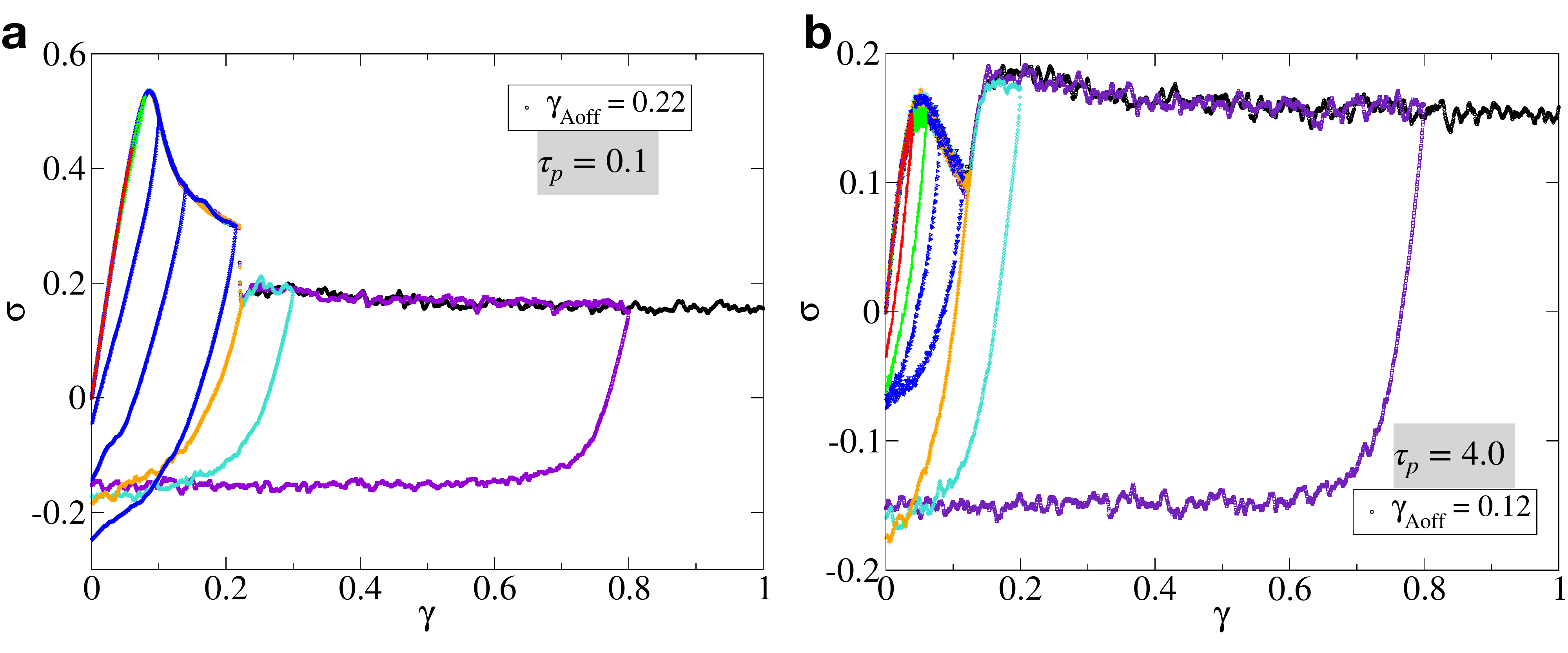}
\caption{In this figure, we show the stress-strain response of the system when the shear direction is reversed in the case where activity is switched off at $\gamma_{Aoff} = 0.22$ and $\gamma_{Aoff} = 0.12$ in (a) and (b) for a persistence time of $\tau_p = 0.1$, and panel (b) to $\tau_p = 4.0$. This demonstrates that, upon deactivating activity, the system immediately exhibits a drop in the residual stress value, similar to that observed in the passive system.}
\label{SS_ActOff}
\end{figure*} 

In Fig.~\ref{SS_ActOff}, Panel (a) and (b) show the effect of switching off activity at $\gamma_{Aoff} = 0.22$ and $\gamma_{Aoff} = 0.12$ for $\tau_p = 0.1$ and $4.0$, respectively. This leads to a transient steady state right after switching off the activity, and the stress subsequently saturates.
We observe that once the activity is switched off, the system stress drops immediately, and its $\sigma(\gamma_0)$ also immediately stops decreasing any further but instead starts saturating to a value. Thus, we show that the presence of activity is important for lower $\sigma(\gamma_0)$ and a gradual decrease in $\sigma(\gamma_0)$. These findings support our results that activity plays a crucial role in determining the mechanical memory of ultrastable glasses by controlling the formation and evolution of multiple shear bands.

\section{Oscillatory Shear at different strain amplitudes}

\begin{figure*}
\includegraphics[width=0.99\textwidth]{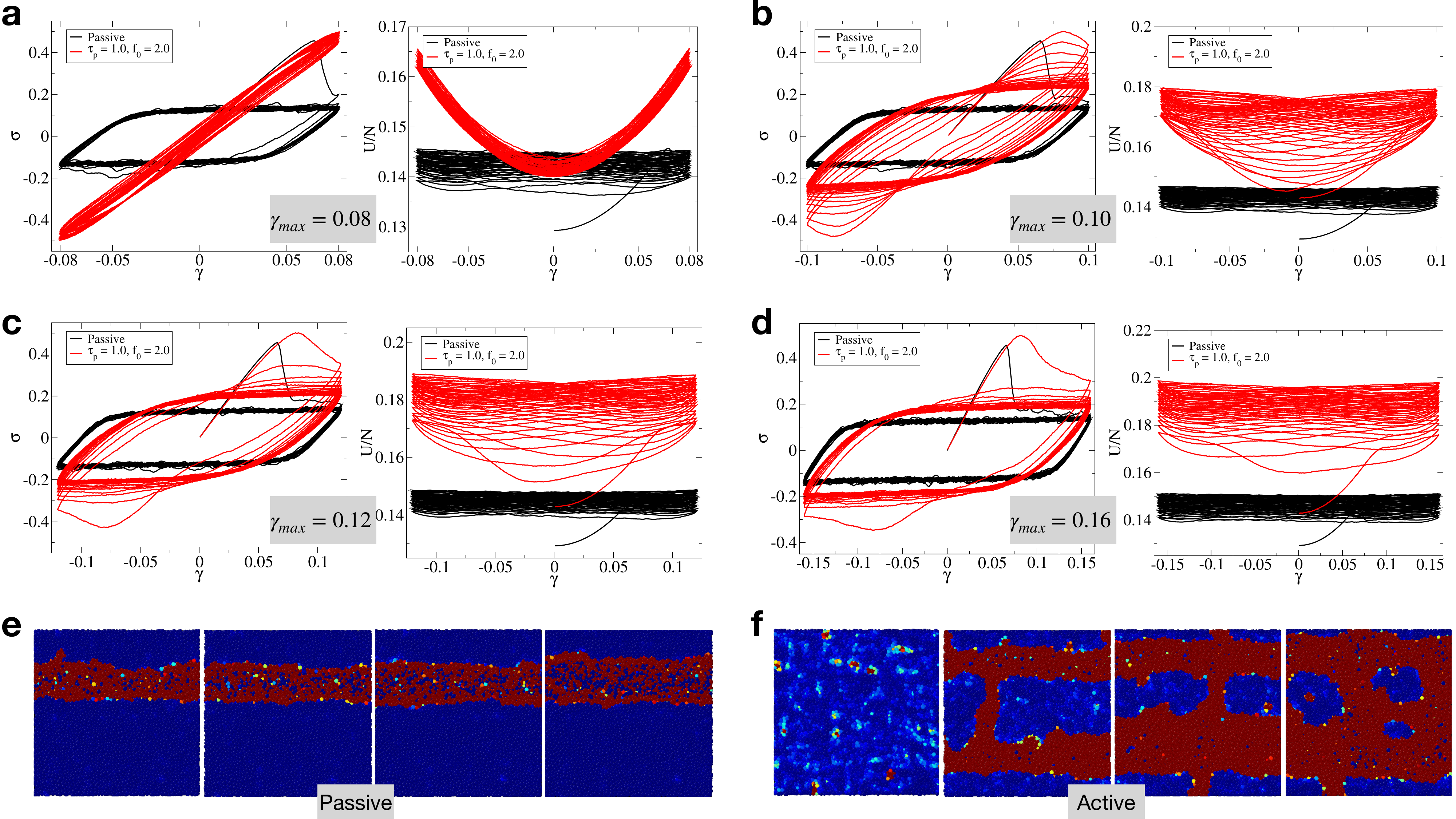}
\caption{This figure illustrates the mechanical response of passive and active particulate systems subjected to cyclic shear deformation. Panels (a-d) compare stress-strain relationships (left) and potential energy per particle (right) across increasing maximum strain amplitudes ($\gamma_{max}$ is equal to $0.08$, $0.10$, $0.12$, and $0.16$, respectively). Black curves represent the passive system response, while red curves show the active system behavior with parameters $\tau_p = 1.0$ and $f_ = 2.0$. The stress-strain hysteresis loops reveal distinct differences in material yielding and energy dissipation between passive and active cases. Panels (e) and (f) display the spatial configuration of shear bands after $25$ deformation cycles in passive and active systems, respectively, highlighting fundamentally different structural reorganization patterns. The active system exhibits more pronounced energy fluctuations and complex shear band morphology compared to its passive counterpart.}
\label{OS_Shear_detailed}
\end{figure*}

The cyclic shear experiments reveal significant differences between passive and active particulate systems under mechanical deformation. Fig.~\ref{OS_Shear_detailed} presents a comprehensive analysis of stress response, energy fluctuations, and structural evolution across varying strain amplitudes. As shown in panels (a-d), both systems exhibit hysteresis behavior under cyclic loading, but with marked differences. The passive system (black curves) demonstrates relatively consistent stress-strain loops across all tested strain amplitudes ($\gamma_{max}$ = $0.08-0.16$), characterized by smooth transitions and moderate energy dissipation. In contrast, the active system (red curves, $\tau_p = 1.0$ and $f_0 = 2.0$) displays pronounced variations in loop morphology and significantly greater stress amplitudes, particularly at intermediate strain values.

The potential energy landscapes (right plots in panels a-d) further highlight these differences. While the passive system maintains relatively stable energy profiles during cycling, the active system experiences substantial energy fluctuations that intensify with increasing strain amplitude. Perhaps most striking are the distinct patterns of shear band formation after extended cycling (panels e and f). The passive system develops system-spanning shear bands, which continue to span with an increasing number of cycles. In contrast, the active system, due to the formation of multiple shear bands, which reshapes in heterogeneous structures, finds it easier to span and reach a steady state with the increasing number of cycles. Additionally, the active system exhibits greater cycle-to-cycle variability that is absent in passive materials. These results demonstrate that activity fundamentally transforms how materials respond to mechanical deformation in ultrastable glasses.

